\documentstyle[epsfig]{aipproc}

\begin{document}

\def\ee{$e^\pm$}
\def\g{$\gamma$}
\def\nh{N_{\rm H}}
\def\af{A_{\rm Fe}}
\def\taut{\tau_{\rm T}}
\def\ginga{{\it Ginga}}
\def\asca{{\it ASCA}}
\def\heao{{\it HEAO-1}}
\def\xte{{\it RXTE}}
\def\rosat{{\it ROSAT}}
\def\ec{E_{\rm c}}
\def\efe{E_{\rm Fe}}
\def\sfe{\sigma_{\rm Fe}}
\def\ife{I_{\rm Fe}}
\def\loc{\ell_{\rm l}}
\def\solm{M$_\odot$}

\title{X-ray and $\gamma$-ray spectra of Cyg X-1 in the soft state}

\author{Marek Gierli\'nski$^*$, Andrzej A. Zdziarski$^\dagger$, Tadayasu 
Dotani$^\ddagger$, Ken Ebisawa$^{\S}$, Keith Jahoda$^{\S}$ and W. Neil 
Johnson$^{\P}$}

\address{$^*$Jagiellonian University Observatory, Cracow, Poland\\
$^{\dagger}$Copernicus Astronomical Center, Warsaw, Poland\\
$^{\ddagger}$Institute of Space and Astronautical Science, Sagamihara, Japan\\
$^{\S}$NASA/GSFC, Greenbelt, USA\\
$^{\P}$E. O. Hulburt Center for Space Research, Naval Research 
Laboratory, Washington, USA
}

\maketitle

\begin{abstract}

We present X-ray/$\gamma$-ray observations of Cyg X-1 in the soft state 
during 1996 May--June. We analyze \asca, \xte\ and OSSE data. The 
spectrum consists of soft X-ray blackbody emission of an optically thick 
accretion disk in the vicinity of a black hole and a power law with an 
energy index $\alpha \sim 1.2$--1.5 extending to at least several 
hundred keV. In the spectra, we find the presence of strong Compton 
reflection, which probably comes from the disk. 

\end{abstract}

\section*{Introduction}

Cyg X-1, the primary black-hole candidate, undergoes transitions between 
two spectral states: the hard (`low') state in which its X-ray spectrum 
is hard ($\alpha \sim 0.6$) and extends up to several hundred keV, and 
the soft (`high') state dominated by a strong soft X-ray emission 
together with a much softer power law ($\alpha \sim 1.5$) tail. Usually, 
Cyg X-1 remains in the hard state. The last transition to the soft state 
began around 1996 May 16 when the soft X-ray flux started to increase. 
The source remained in the soft state until about August 11 (Zhang et 
al.\ 1997).

\section*{The data}

In this paper, we analyze three groups of observations of Cyg X-1 in 1996. An 
\xte\/ observation on May 23 shows the object still undergoing a transition 
between the states (Fig.\ 1). The remaining data sets are from a simultaneous 
\asca/\xte\/ observation on May 30 and from six simultaneous \xte/OSSE 
observations on June 17-18, when the object was in the soft state. 

The \xte\/ data come from the public archive. A 2\% systematic error has been 
added to each PCA channel to represent calibration uncertainties. Since 
dead-time effects of the HEXTE clusters are not yet fully understood, we 
allowed a free relative normalization of the HEXTE data with respect to the 
PCA data. 

The {\it CGRO}/OSSE observation in the soft state on June 14--25 (from 50 keV 
to 1 MeV) overlaps with the \xte\ observations on June 17 and 18. We 
have extracted six OSSE data sets near-simultaneous with the \xte\ 
observations. In order to get better statistics, we have increased each OSSE 
data interval to include 30 minutes on either side of the corresponding \xte\ 
interval. The spectrum is shown in Fig.\ 1, which also shows a spectrum from 
the hard state for comparison. 

\asca\/ observed Cyg X-1 from May 30 5:30 (UT) through May 31 3:20 
(Dotani at al.\ 1997, hereafter D97). The GIS observation was made in 
the standard PH mode. The SIS data suffer from heavy photon pile-up and 
thus are not usable. We have selected 1488 live seconds of the \asca\/ 
data near-simultaneous with the corresponding \xte\ observation.

\begin{figure}[t!]
\centerline{\epsfig{file=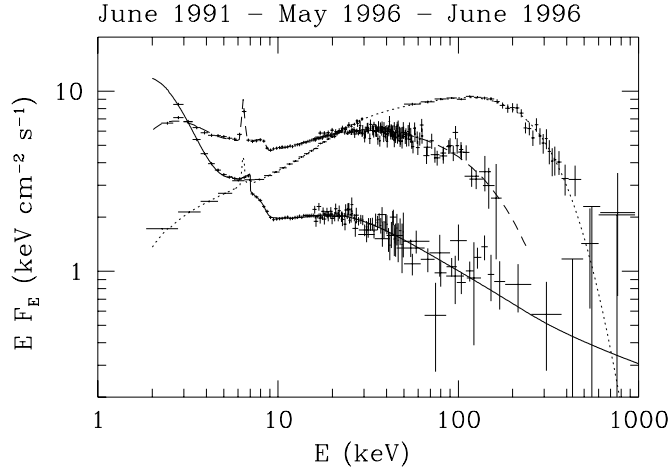,width=3.5in}}
\vspace{10pt}
\caption{The spectral states of Cyg X-1. The solid curve is a fit to the 
soft state observed by \xte\ and OSSE on 1996 June 17. The dashed curve 
is a fit to an intermediate state observed by \xte\ on 1996 May 23. For 
comparison, we also show a \ginga/OSSE spectrum in the hard state (on 
1991 June 6) with the model represented by the dotted curve (see 
Gierli\'nski at al.\ 1997). The data have been rebinned for clarity.} 
\label{transitions} 
\end{figure}

\section*{Results}

First, we fit the simultaneous \xte/OSSE soft-state data of June 17--18. 
At soft X-rays, the spectra are dominated by a black-body component. The 
OSSE data show the emission extending up to at least 800 keV. Our basic 
model consists of a soft X-ray blackbody disk model and a power law. The 
power-law energy index, $\alpha$, varies between $\sim 1.3$ and $\sim 
1.5$. We do not observe any high-energy cutoff in the power law, which 
suggests a non-thermal origin of the emission. We note that the 
power-law component comes probably from Comptonization of the soft 
photons and should be therefore cut off at low energies. However, this 
cutoff would occur only well below 1 keV and its neglect in the model 
does not affect our conclusions. 

Between $\sim 10$--200 keV, the observed spectrum systematically departs
from the power law and forms a smooth hump. There is also a broad 
absorption feature above 7 keV, which can be attributed to an Fe 
K$\alpha$ absorption edge. Considering both effects, we conclude that 
Compton reflection from cold matter takes place in the soft state of Cyg 
X-1. The covering factor of the reflector, $\Omega/2\pi$, varies in the 
range 0.6-0.7. The reflector can be identified with the optically thick 
disk also responsible for the soft blackbody.


The iron edge is smeared, which we attribute to Doppler and 
strong-gravity effects in the vicinity of the black hole. Therefore, we 
consider models in the Schwartzschild metric with the accretion disk 
inclined at angle 35$^\circ$ and reflection taking place between $R_{in} 
= 3 R_g$ and $R_{out} = 15 R_g$ (where $R_g = 2GM/c^2$). In this paper, 
we use a model of angle-dependent Compton reflection (Magdziarz \& 
Zdziarski 1995) convolved with the relativistic line profile (Fabian at 
al.\ 1989). We note that this is only an approximation of the real disk 
reflection and does not take into account all angular effects near the 
black hole. 

\begin{figure}[t!]
\centerline{\epsfig{file=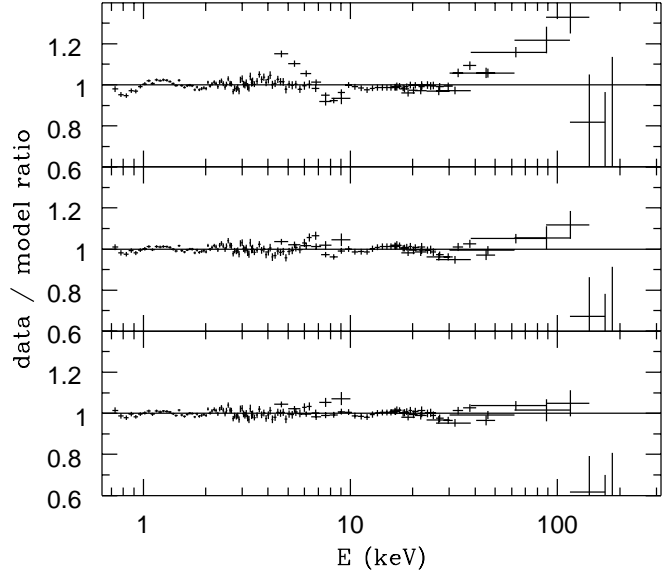,width=3.5in}}
\vspace{10pt}
\caption{The data-to-model ratios for the 1996 May 30 observation. The upper 
panel corresponds to the model including power law, reflection and disk 
emission, the middle panel includes a Comptonization tail to the disk 
emission, and the lower panel includes the tail and an Fe line.} 
\label{ratios} \end{figure}

\begin{figure}[t!]
\centerline{\epsfig{file=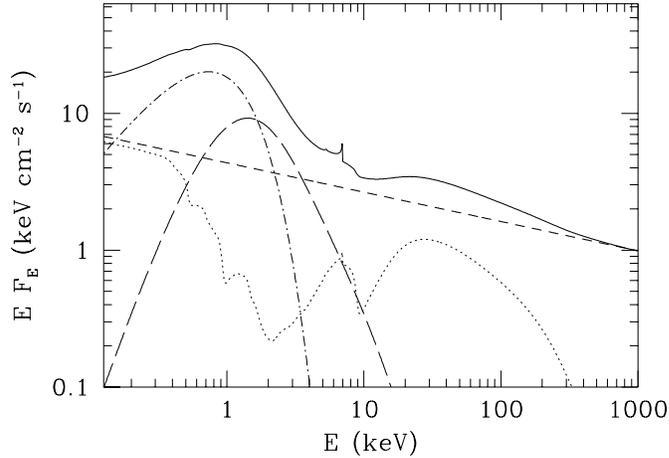,width=3.5in}}
\vspace{10pt}
\caption{The model spectrum of Cyg X-1 in the soft state based on the 
simultaneous \asca/\xte\ observations of 1996 May 30. The short-dashed, 
dotted, dash-dotted, and long-dashed  curves show the power law, the Compton 
reflection, the blackbody-disk emission, and the thermal Comptonization 
components, respectively. Solid line is the sum of all components. In order to 
show the details of the soft X-ray emission, the interstellar absorption was 
removed from the model.} \label{maymodel} \end{figure} 

We then study the \asca/\xte\ data of May 30. We apply a conservative 
lower limit of 4 keV to the PCA data, and use the \asca\ data in the 
range of 0.7--10 keV. First, we use a blackbody disk emission model 
taking into account general relativity (Hanawa 1989) together with a 
power law with relativistic reflection (as for the June data above). 
This, however, yields an unacceptable $\chi^2$ of about 1100/577 d.o.f.\ 
(see the residuals in the upper panel of Fig.\ 2). The fit can be 
significantly improved by an additional weak high-energy tail on top of 
the disk spectrum. We model this tail as due to thermal Comptonization 
of a 300 eV blackbody in a plasma with $kT \approx 5$ keV and $\tau 
\approx 3$ (shown by the long-dashed curve on Figure \ref{maymodel}). 
The resulting $\chi^2$ is 680/575 d.o.f. The additional component can be 
interpreted as Compton radiation of an intermediate layer between the 
optically thick disk and an optically thin corona. We note that a 
similar reduction of $\chi^2$ can be (instead of adding the 
Comptonization tail) obtained by breaking the power law to a softer one 
below a few keV. A similar softening of the power law at low energies 
was observed by \asca\/ in the hard state (Ebisawa et al.\ 1996). 

The fit can be further improved by adding an Fe K$\alpha$ line. For that,
we use a relativistic disk line (Fabian at al.\ 1989) with the same 
parameters as for the Compton reflection. The obtained $\chi^2$ is 641/573 
d.o.f., i.e., the presence of line is statistically significant at a very high 
confidence level. Figure \ref{ratios} shows the residuals corresponding to the 
above models. The model spectrum is presented in Fig.\ \ref{maymodel}. 

The spectrum is absorbed by $N_H = 0.47\pm0.01$. The fitted power-law 
index is $\alpha$ = 1.23$\pm 0.03$. We assume no cutoff in the power law 
(as implied by the June 14--25 OSSE data). The covering factor of the 
reflector is $\Omega/2\pi= 0.55\pm 0.1$, and the reflecting medium is 
ionized with ionization parameter $\xi \equiv L / nr^2 = 
430^{+160}_{-130}$ erg cm s$^{-1}$, corresponding to Fe {\sc xxi-xxiv} 
as the most abundant species. 


The best fit value of the rest-frame line energy, $E_{\rm Fe} = 
6.54^{+0.11}_{-0.15}$ keV, strongly depends on the assumed disk 
inclination angle. Since the Auger resonant destruction strongly 
suppresses K$\alpha$ emission following photoionization of Fe\,{\sc 
xvii-xxiii}, we should expect fluorescence from lithium-like Fe {\sc 
xxiv} at $\sim$\,6.7 keV. On the basis of our model, this line energy is 
consistent with the disk inclination angle of 30$^\circ$. 
Notwithstanding, we stress that due to approximate character of the 
model and uncertainties in the PCA response matrices, the line 
parameters obtained here might be inaccurate. 

If we use the same blackbody disk model as D97 assuming $T_{\rm 
col}/T_{\rm eff}$ = 1.7, we obtain a black hole mass of 21$\pm 2$ 
M$_{\odot}$ and an accretion rate of $6.5\pm 0.4\times10^{17}$ g 
s$^{-1}$. That mass is significantly higher than $12^{+3}_{-1}$ \solm\ 
found by D97. The discrepancy is explained by influence of the 
additional Comptonization component used here. On the other hand, we 
obtain $M_x \approx$ 16 \solm\ by assuming $T_{\rm col}/T_{\rm eff}$ = 
1.5. We stress that the ratio of $T_{\rm col}/T_{\rm eff}$ is very 
uncertain due to theoretical difficulties. In particular, the standard 
Shakura-Sunyaev solution is unstable at $\dot m \equiv \dot Mc^2/L_{\rm 
Edd} \sim 1$. In Cyg X-1 the total disk luminosity in the soft state is 
$L_{\rm disk} \approx 5 \times 10^{37}$ erg s$^{- 1}$ (about 5 times 
higher than in the hard state), which corresponds to $\sim 0.05 L_{\rm 
Edd}$. Assuming the emission efficiency of 0.08, we find $\dot m \sim 
1$. 




\begin{references}


\bibitem{dotani97}Dotani T., et al.\ {\it ApJL}, in press (1997).

\bibitem{ebisawa96}Ebisawa K., at al.\ {\it ApJ} {\bf 467}, 419 (1996).

\bibitem{fabian89}Fabian A. C., at al.\ {\it MNRAS} {\bf 238} 729 
(1989).

\bibitem{gierlinski97}Gierli\'nski M., at al.\ {\it MNRAS}, in press 
(1997).

\bibitem{magdziarz95}Magdziarz P. and Zdziarski A., {\it MNRAS} {\bf 
273}, 837 (1995).

\bibitem{zhang97}Zhang S. N., at al.\ {\it ApJ} {\bf 477}, L95 (1997).

\end{references}
\end{document}